\newif\ifsingle

\newif\ifFullVersion
\FullVersiontrue 

\ifsingle
\documentclass[12pt,draftclsnofoot, onecolumn]{IEEEtran}		
\else		
\documentclass[10pt,final, twocolumn]{IEEEtran}
\fi

\usepackage{times}
\usepackage{amsmath,dsfont}
\usepackage{amssymb,amsthm}
\usepackage{epsfig,verbatim}
\usepackage{subcaption}
\usepackage{setspace}
\usepackage{color}
\usepackage{cite}
\usepackage{epstopdf}
\usepackage{graphics}
\usepackage{accents}
\usepackage{acronym}
\usepackage{booktabs}
\usepackage{mathtools} 
\usepackage{enumitem}
\usepackage{multirow}
\usepackage{dblfloatfix}
\usepackage[bookmarks,colorlinks]{hyperref}
\usepackage{amsmath}
\usepackage{gensymb}
\usepackage[flushleft]{threeparttable}
\usepackage{algpseudocode}

\usepackage[linesnumbered,ruled,lined]{algorithm2e}

\SetKwInput{KwData}{\textbf{Init}} 
\let\oldnl\nl
\newcommand{\nonl}{\renewcommand{\nl}{\let\nl\oldnl}}

\newcommand{\myVec}[1]{{\boldsymbol{#1}}}

\newcommand{\mySet}[1]{\mathcal{#1}}
	
\makeatletter
\newcommand{\removelatexerror}{\let\@latex@error\@gobble}
\makeatother

\ifsingle

\else

\fi 

\acrodef{adc}[ADC]{analog-to-digital convertor}
\acrodef{cs}[CS]{compressed sensing}
\acrodef{dtft}[DTFT]{discrete-time Fourier transform}
\acrodef{dnn}[DNN]{deep neural network} 
\acrodef{csi}[CSI]{channel state information}
\acrodef{bpsk}[BPSK]{binary phase shift keying}
\acrodef{map}[MAP]{maximum a-posteriori probability}
\acrodef{snr}[SNR]{signal-to-noise ratio}
\acrodef{bs}[BS]{base station} 
\acrodef{iot}[IOT]{Interent of Things}
\acrodef{mimo}[MIMO]{multiple-input multiple-output}
\acrodef{siso}[SISO]{single-input single-output}
\acrodef{mse}[MSE]{mean-squared error}
\acrodef{pdf}[PDF]{probability density function}
\acrodef{rv}[RV]{random variable}
\acrodef{mle}[ML]{maximum likelihood}
\acrodef{ml}[ML]{machine learning}
\acrodef{fec}[FEC]{forward error correction}
\acrodef{rs}[RS]{Reed-Solomon}
\acrodef{lti}[LTI]{linear time-invariant}
\acrodef{wss}[WSS]{wide-sense stationary}
\acrodef{psd}[PSD]{power spectral density}
\acrodef{ser}[SER]{symbol error rate} 
\acrodef{ber}[BER]{bit error rate} 
\acrodef{fer}[FER]{frame error rate} 
\acrodef{gd}[GD]{gradient descent}
\acrodef{sgd}[SGD]{stochastic gradient descent} 
\acrodef{isi}[ISI]{intersymbol interference}  
\acrodef{awgn}[AWGN]{additive  white  Gaussian noise} 
\acrodef{ut}[UT]{user terminal} 
\acrodef{mmw}[mmWave]{millimeter wave}
\acrodef{noma}[NOMA]{non-orthogonal multiple access}
\acrodef{mac}[MAC]{mulitple access channel}
\acrodef{fl}[FL]{Federated learning}
\acrodef{lstm}[LSTM]{long short-term memory}
\acrodef{maml}[MAML]{model-agnostic meta-learning}
\acrodef{sic}[SIC]{soft interference cancellation}
\acrodef{pmf}[PMF]{probability mass function}
\acrodef{sota}[SOTA]{state of the art}
\acrodef{ecc}[ECC]{error-correction code}
\acrodef{crc}[CRC]{cyclic redundancy check}
\acrodef{scl}[SCL]{successive cancellation list}
\acrodef{sc}[SC]{successive cancellation}
\acrodef{bp}[BP]{belief propagation}
\acrodef{wbp}[WBP]{weighted belief propagation}
\acrodef{urllc}[URLLC]{ultra-reliable low latency communications}
\acrodef{mmtc}[mMTC]{massive machine-type communications}
\acrodef{llr}[LLR]{log-likelihood ratio}
\acrodef{em}[EM]{expectation maximization}
\acrodef{msb}[MSB]{most significant bit}
\acrodef{lsb}[LSB]{least significant bit}
\acrodef{flops}[FLOPS]{floating-point operations per second}
\acrodef{nn}[NN]{neural network}
\acrodef{bce}[BCE]{Binary Cross Entropy}

\IEEEoverridecommandlockouts 

\title{CRC-Aided Learned Ensembles of Belief-Propagation Polar Decoders}
\author{  
	\IEEEauthorblockN{Tomer Raviv, Alon Goldman, Ofek Vayner, Yair Be'ery, and Nir Shlezinger
	} 
	\thanks{This project has received funding from the Israeli 5G-WIN consertium.
    A. Goldman, O. Vayner, T. Raviv and N. Shlezinger are with the School of ECE, Ben-Gurion University of the Negev, Beer-Sheva, Israel (e-mail: alon.goldmann@gmail.com, ofekvay@post.bgu.ac.il, tomerraviv95@gmail.com, nirshl@bgu.ac.il).
    Y. Be'ery is with the School of Electrical Engineering, Tel-Aviv University, Israel (e-mail: ybeery@eng.tau.ac.il). 
    }

	\vspace{-1.0cm}
	
}
\vspace{-0.75cm}

\begin{document}
	
	\maketitle

	\pagestyle{plain}
	\thispagestyle{plain}
	\begin{abstract} 
Polar codes  have  promising error-correction capabilities. Yet, decoding polar codes is often challenging, particularly with large blocks, with recently proposed decoders  based on list-decoding or neural-decoding. The former applies multiple decoders or the same decoder multiple times with some redundancy, while the latter family utilizes emerging deep learning schemes to learn to decode from data.
In this work we introduce a novel polar decoder that combines the list-decoding with neural-decoding, by forming an ensemble of multiple weighted belief-propagation (WBP)  decoders, each trained to decode different data. We employ the cyclic-redundancy check (CRC) code  as a proxy for combining the ensemble decoders and selecting the most-likely decoded word after inference, while facilitating real-time decoding. We evaluate our scheme over a wide range of polar codes lengths, empirically showing that gains of around $0.25$dB in frame-error rate could be achieved. Moreover, we provide complexity and latency analysis, showing that the number of operations required approaches  that of a single BP decoder at high signal-to-noise ratios.
\end{abstract}
	\vspace{-0.4cm}
	\section{Introduction}
\vspace{-0.1cm} 
	
Wireless data traffic have grown exponentially over recent years with no foreseen saturation. To keep pace with connectivity requirements, one must carefully attend available resources with respect to three essential measures: reliability, latency and complexity. As \acp{ecc} are well renowned as means to boost reliability, the research of practical schemes is crucial to meet demands.

Polar codes \cite{arikan2009channel} are novel \acp{ecc} incorporated in 5G frameworks. Their polarization principle dictates that splitting and combining the different channels allows one to form a set of "good" and "bad" channels, where the polar encoder transmits the information part only in the set of "good" channels. 
While such encoding can be theoretically guaranteed to approach the symmetric capacity for any given binary-input memoryless channel when using \ac{mle} decoding, such decoders are associated with increased complexity; practical polar decoders are typically associated with a tradeoff between performance and complexity. 
A common polar decoding scheme uses \ac{sc} \cite{alamdar2011simplified}. \ac{sc} decoders  recursively reduce the problem size by decoding a single bit in each iteration. While \ac{sc} is of a relatively low complexity, its sequential operation induces notable latency, and its performance is typically within   a significant performance gap from that of  \ac{mle} decoding.  
To close this gap, the \ac{scl} algorithm \cite{tal2015list} was proposed. \ac{scl} decodes  a number of different paths concurrently, using a \ac{crc} precoding as a measure to eliminate non-likely paths and select the most likely one. Though the performance of the above \ac{scl} is close to \ac{mle}, it incurs high latency and complex hardware which is unfit for low latency applications. 
An alternative polar decoding approach utilizes  \ac{bp}, which is parallelizable and  computing-efficient, bridging the latency and complexity gap. Yet, \ac{bp} is often outperformed by \ac{scl}.

To exploit the full potential  of the \ac{bp}, different improvements were suggested in the decoding literature.  
The first set of works includes \ac{bp}-list decoder, which is composed of multiple parallel independent \ac{bp} decoders that output a list of candidate decoded words. All candidates in the list are ranked by some metric, and the single most suitable one is output. One  important principle that encourages different candidates is the principle of diversity: how diverse is every decoder to the combination. To achieve diversity among the different decoders, one can employ different measures, such as adding synthetic noise realizations \cite{arli2019noise} or applying random permutations \cite{elkelesh2018belief}. However, if the noise does not represent the channel well, or the number of  permutations is large, one has to resort to  heuristics \cite{geiselhart2020crc}, typically with 
 suboptimal performance.

A second set of works focused on integrating deep learning into the basic decoder design, as a form of model-based deep learning~\cite{shlezinger2022model}. In particular, the deep unfolding methodology was utilized in \cite{nachmani2018deep} to leverage data to improve \ac{bp} decoding with a fixed number of iterations, and was adapted to \ac{bp}-based polar decoding in \cite{xu2017improved}. 
%
Soon after, more improvements to the basic \ac{dnn}-aided \ac{bp} decoder followed: These include highly-parameterized hypernetworks that were proposed in \cite{nachmani2020gated} to decode polar codes by employing a new formalization of \ac{bp} decoding, and \cite{choukroun2022error} which introduced a specialized transformer architecture. Finally, to select permutations that mitigate the detrimental effects of cycles to improve performance, \cite{caciularu2020perm2vec} presented a data-driven framework for permutation selection. Though these solutions improve performance, they lose the complexity merits of \ac{bp}, particularly for large blocks, limiting their suitability for low-complexity applications.



In this work we propose a polar decoding scheme that combines list decoding concepts,  exploiting redundancy to boost performance, with low-complexity deep-learning-aided \ac{bp} decoding. To that end, we adapt the data-driven ensembles scheme presented in \cite{raviv2020data} to the case of polar codes, while basing the decoders and the selection rule on 5G compliant \ac{crc} codes. The ensemble is composed of learnable decoders, each exploiting the polar code factor graph for decoding, while harnessing the
%
\ac{crc} code for gating and aggregation. The main intuition behind this divide-and-conquer approach is that combination of multiple \textit{diverse} trainable decoders is expected to perform better than each individual decoder.

Our main contributions are summarized as follows: 
\begin{itemize}
    \item \textbf{Polar \ac{bp} ensemble:} We formulate a dedicated ensemble framework for concatenated polar and \ac{crc} codes. The ensemble results in low  latency and compute, and is easily applicable to medium-to-high lengths codes, where other \ac{dnn}-aided decoders \cite{nachmani2020gated,choukroun2022error}  struggle.
    \item \textbf{Extensive experimentation:} We extensively evaluate our decoder for various code lengths, validating that our gains are consistent in all regimes. The latency of the scheme is analyzed and is shown to only induce minor increase compared with standard \ac{bp}.
\end{itemize}

 The rest of this paper is organized as follows:  Section~\ref{sec:system_model} details the system model, while Section~\ref{sec:polar_ensembles} formulates the framework. The experimental results and concluding remarks are detailed in Section~\ref{sec:numerical_evaluations} and Section~\ref{sec:conclusion}, respectively.

Throughout the paper, we use calligraphic letters  for sets, e.g., $\mySet{X}$, and boldface letters for vectors, e.g., ${\myVec{x}}$; $({\myVec{x}})_i$ denotes
the $i$th element of ${\myVec{x}}$; for bit vectors, $i=0$  is the \ac{msb}. 
We denote by $\boldsymbol{0}$ and $\boldsymbol{1}$ the all-zeros and all-ones vector, respectively, and by $(\cdot)^T$ the transpose.
	\vspace{-0.2cm}
	\section{System Model}
\label{sec:system_model}
\vspace{-0.1cm}
In this section we formulate the problem and review some preliminaries. We first describe encoding procedure in Subsection~\ref{subsec:crc_codes}. Following is the \ac{bp} polar decoder, recalled in Subsection~\ref{subsec:wbp}, and the problem formulation  in Subsection~\ref{subsec:problem_formulation}. 

\vspace{-0.2cm}
\subsection{Encoding Scheme}
\label{subsec:crc_codes}
\vspace{-0.1cm}  
We consider a point-to-point communication scenario where the transmitter encodes via \ac{crc} and polar coding.

\smallskip
{\bf \ac{crc} Codes:} 
The first steps encodes the message using linear binary cyclic error detection codes based on \ac{crc}. 
\ac{crc}  encoding maps a sequence of $N_m$ information bits $\boldsymbol{m}=[m_0, ..., m_{N_m-1}]$ into an $N_u$ bits codeword $\boldsymbol{u}$ by adding $N_u-N_m$ parity bits $\boldsymbol{r}=[r_0, ..., r_{N_u-N_m-1}]$. To calculate these parity bits, one has to calculate division of the information polynomial by the \ac{crc} generator polynomial. The polynomial representation of the information word is  $\boldsymbol{m}(x)=m_0+m_1x+...+m_{N_m-1}x^{N_m-1}$, while $\boldsymbol{g}_{CRC}(x)$ is the generator polynomial. For example, for the length 11 CRC code used in our numerical study and specified for control in 5G \cite{baicheva2021some}, the generator is $\boldsymbol{g}_{CRC}(x)=x^{11}+x^{10}+x^{9}+x^{5}+1$. 

The polynomial representation  of the codeword, denoted $\boldsymbol{u}(x)$, is given by the multiplication of $\boldsymbol{g}_{CRC}(x)$ and $\boldsymbol{m}(x)$. Thus, every candidate word $\hat{\boldsymbol{u}}\in\{0,1\}^{N_u}$ is valid if and only if its polynomial $\hat{\boldsymbol{u}}(x)$ is divisible by $\boldsymbol{g}(x)$, meaning that the polynomial division results in the zeros vector. Denoting the generator matrix of the \ac{crc} code by $\boldsymbol{G}_{CRC}$ and its parity-check matrix by $\boldsymbol{H}_{CRC}$, this yields the encoding equation:
\begin{equation}
\label{eqn:crc_encoding}
	\boldsymbol{u} = \boldsymbol{G}_{CRC}^T\boldsymbol{m}
\end{equation}
and the parity-check validation rule:
\begin{equation}
\label{eqn:parity-check}
	\boldsymbol{r} = \boldsymbol{H}_{CRC} \boldsymbol{\hat{u}}
\end{equation}
where $\boldsymbol{r} = \boldsymbol{0}$ for a valid \ac{crc} codeword.

\begin{figure}
    \centering
    \includegraphics[width=0.8\columnwidth]{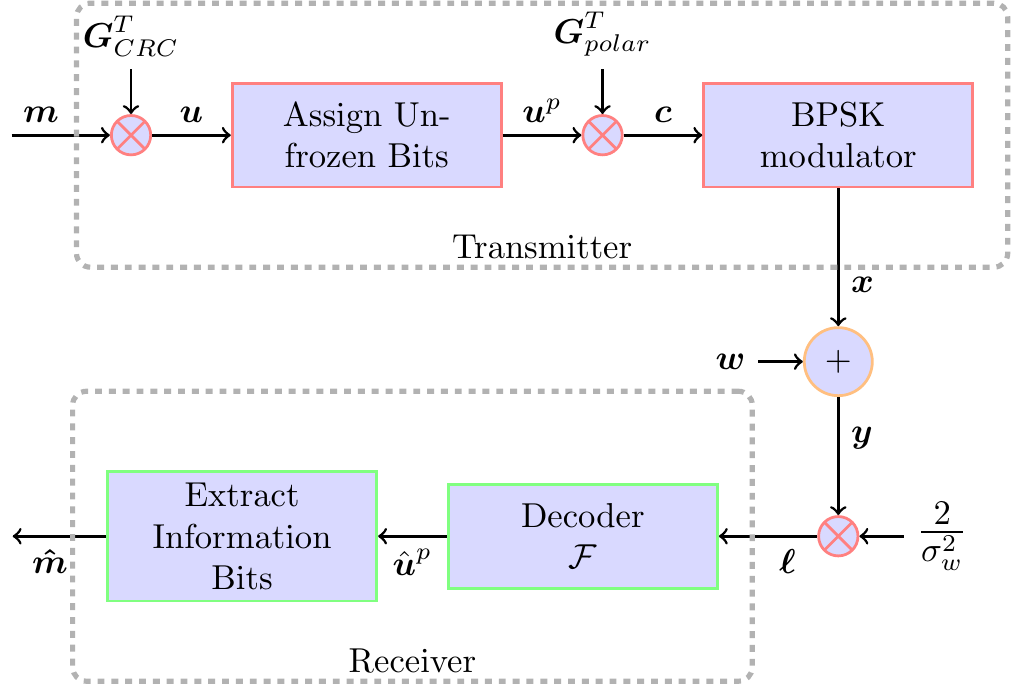}
    \caption{System diagram}
    \label{fig:system_diagram}
\end{figure}

\smallskip
{\bf Polar Codes:}
The polar transformation of size $(N_c,N_u)$ polar code is defined by two main steps. First, a \ac{crc} encoded codeword $\boldsymbol{u}\in\{0,1\}^{N_u}$ is mapped to the zeros-padded $\boldsymbol{u}^{p}\in\{0,1\}^{N_c}$ by placing the $N_u$ bits of $\boldsymbol{u}$ in the reliable positions of $\boldsymbol{u}^{p}$, while zeros are placed in the unreliable frozen positions. The reliable positions are set, e.g., by Bhattacharyya parameter's construction. Then, the polar encoder applies the polarization matrix $\boldsymbol{G}_{polar} = \boldsymbol{F}^{\otimes{n_c}}$, where $
\boldsymbol{F}=
\big[\begin{smallmatrix} 
1 & 0\\ 1 & 1 
\end{smallmatrix}
\big]$ is the basic polarization kernel, $N_c$ is the code length, $n_c=\log_2 N_c$ and $\boldsymbol{F}^{\otimes{n}}$ is the $n^\text{th}$ Kronecker power of the kernel. The desired codeword is given by $\boldsymbol{c} =\boldsymbol{G}_{polar}^{T} \boldsymbol{u}^{p}$.

\smallskip
{\bf Transmission Model:}
After encoding, the codeword $\boldsymbol{c}$ is BPSK-modulated into  $\boldsymbol{x}\in\{-1,1\}^{N_c}$  ($0\mapsto1$, $1\mapsto-1$), and $\boldsymbol{x}$ is transmitted through a channel with \ac{awgn} $\boldsymbol{w}\sim\mathcal{N}(\boldsymbol{0},\sigma_w^2 \boldsymbol{I})$. The received word  denoted  $\boldsymbol{y}$, is used to compute the \ac{llr} word $\boldsymbol{\ell}= \frac{2\boldsymbol{y}}{\sigma_w^2}$. The \ac{llr} values are approximated based on the bits i.i.d. assumption and the Gaussian prior. The decoder is represented by a function $\mathcal{F}:\mathbb{R}^{N_c}\rightarrow\{0,1\}^{N_c}$ that outputs the estimated zeros-padded detection codeword $\hat{\boldsymbol{u}}^{p}$, from which the codeword $\hat{\boldsymbol{u}}$ is extracted and the message $\hat{\boldsymbol{m}}$ is recovered. The considered system model 
is  depicted in Fig.~\ref{fig:system_diagram}.

\vspace{-0.2cm}
\subsection{Weighted Belief Propagation}
\label{subsec:wbp}
\vspace{-0.1cm}

\Ac{bp} efficiently calculates the marginal probabilities of nodes in a graph. 
 A graphical representation used in channel coding is the Tanner graph  of  linear error-correcting codes. A Tanner graph is a bipartite graph with two types of nodes: variable nodes and check nodes. The variable nodes represent the bits of the encoded message, and the check nodes represent the constraints 
 that the encoded message must satisfy. 

In the presence of loops in the graph, the \ac{bp} decoder $\mathcal{F}_{BP}$ is generally suboptimal. Its performance can be improved by weighting the messages via a learnable variant of the standard \ac{bp}, called \ac{wbp} \cite{nachmani2018deep}. In \ac{wbp}, the messages passed between nodes in the graphical model are weighted based on the relative reliability of the information they represent. This was empirically shown to facilitate convergence more quickly and accurately. The \ac{wbp} procedure can be specialized to the polar factor graph formulation \cite{xu2017improved} as is detailed in Algorithm \ref{alg:wbp}. There, $f(x,y)= \ln{\frac{1+xy}{xy}}$ while node $(i,j)$ denotes the $j^\text{th}$ input at the $i^\text{th}$ stage. 

The weights $\boldsymbol{\gamma}$ of \ac{wbp} can be tuned from data. Here, one has access to  a dataset $\mathcal{D}$ composed of $K$ tuples of the form $(\boldsymbol{\ell},\boldsymbol{u}^{p})$. The \ac{wbp} decoder $\mathcal{F}_{WBP}$ is applied to $\boldsymbol{\ell}$ in each tuple, outputting the soft output $\boldsymbol{L}^{(T)}_{1}$ and is thereafter trained to minimize the \ac{bce} loss:
\begin{align*}
    \mathcal{L}(\boldsymbol{u}^{p},\boldsymbol{L}^{(T)}_{1}) = \frac{-1}{N_c}\sum_{j=1}^{N_c} [& u^{p}_j \log{L^{(T)}_{1,j}} + 
    (1\!-\! u^{p}_j) \log{(1\! -\! L^{(T)}_{1,j})}].
\end{align*}

\begin{figure}
\removelatexerror
  \begin{algorithm}[H]
    \caption{\ac{wbp} Inference}
    \label{alg:wbp}
    \SetAlgoLined
    \SetKwInOut{Input}{Input}
    \SetKwInOut{Output}{Output}
    \SetKwProg{TheWBPAlgorithm}{\ac{wbp} Inference}{}{}
    \Input{\ac{llr} word $\boldsymbol{\ell}$
    \newline Number of iterations $T$ \newline Neural weights $\boldsymbol{\gamma}$} 
    \Output{Decoded zeros-padded word $\hat{\boldsymbol{u}}^{p}$}    \TheWBPAlgorithm{$(\boldsymbol{\ell},T,\boldsymbol{\gamma})$}{
    Initialize the right messages: 
    $R_{1,j}^{(1)} = 
    \begin{cases}
		0, & \text{if $j$ is a reliable position}\\
        \infty, & \text{otherwise}
    \end{cases}$\\ \label{line:right_init}
    Initialize the left ones: 
    $L_{n_c+1,j}^{(1)} = \ell_j$\\ \label{line:left_init}
    
         \For{$t\in\{1,...,T\}$}{ 
            Calculate left updates:\\
           $L_{i,j}^{(t)} = \gamma^L_{i,j} f(L_{i+1,2j-1}^{(t)},L_{i+1,2j}^{(t)} + R_{i,j+N_c/2}^{(t)})$\\ \label{line:left_update1}
           $L_{i,j\!+\!N_c/2}^{(t)} = \gamma^L_{i,j\!+\!N_c/2} f(R_{i,j}^{(t)},L_{i\!+\! 1,2j\!- \!1}^{(t)}) \!+\! L_{i\!+\! 1,2j}^{(t)}$\\ \label{line:left_update2}
           
            Calculate right updates:\\
            $R_{i\!+\! 1,2j\!-\! 1}^{(t)} = \gamma^R_{i\!+\! 1,2j\! - \!1} f(R_{i,j}^{(t)},L_{i\! +\! 1,2j}^{(t\!-\! 1)} \!+\! R_{i,j \!+ \! N_c/2}^{(t)})$\\ \label{line:right_update1}
            $R_{i\! +\! 1,2j}^{(t)} = \gamma^R_{i\! +\! 1,2j} f(R_{i,j}^{(t)},L_{i\! + \!1,2j\! -\! 1}^{(t\! -\! 1)}) \!+\! R_{i,j\!+\! N_c/2}^{(t)}$\\ \label{line:right_update2}
        }
        Set hard-decision bits via:
        $\hat{u}^{p}_{j} = 
    \begin{cases}
		1, & \text{if $L^{(T)}_{1,j} \leq 0$}\\
        0, & \text{otherwise}
    \end{cases}$\\ \label{line:hard_decision}
        \KwRet{$\hat{\boldsymbol{u}}^{p}$}
  }
  \end{algorithm}
\end{figure}

\vspace{-0.2cm}
\subsection{Problem Formulation}
\label{subsec:problem_formulation}
\vspace{-0.1cm} 
The current \ac{wbp} decoding approach utilizes the same $\boldsymbol{\gamma}$ to decode every noisy word, without considering the individual characteristics of each word. This fails to account for the inherent variability in the noise patterns affecting different words, leading to suboptimal performance. Our goal is  to design a data-aided polar decoder, which preserves the  applicability of \ac{wbp} to large blocks, as well as its low latency and complexity, while improving upon it in \ac{fer}
\begin{equation}
\label{eqn:optimization}
	{\rm FER} = 
 \Pr\left( \hat{\boldsymbol{m}}\neq \boldsymbol{m} \right).
\end{equation}

Due to the complexity and latency constraints, we propose to employ multiple low-complexity models that run in parallel therefore limiting the increase in latency. Due to the inherent low complexity of the polar \ac{wbp} and the slight complexity increase due to the neural weights, we  employ multiple such models, formulating the super-model as an ensemble.


	\vspace{-0.2cm}
	\section{CRC-aided Learned  Decoders Ensemble}
	\label{sec:polar_ensembles}
	\vspace{-0.1cm}
	In this section we present our proposed polar decoding algorithm, which combines ensemble of trainable \ac{wbp} decoders with \ac{crc}-aided selection criteria for boosting diversity. Our design is inspired by the established  gains of ensemble-decoding based on WBP  for linear codes shown in \cite{raviv2020data}. The gains of ensemble models highly depend on the ability to produce {\em diverse} agents; the decoder of \cite{raviv2020data} relied on estimating the Hamming distance of the processed block, with no additional external data, as a measure for diversifying and selecting agents. Our proposed decoder leverages  ensemble-decoding, while tailoring its \ac{wbp} agents for polar codes, and using \ac{crc} codes  as external informative domain-oriented measure for diversifying and selecting agents. We next present the resulting decoder operation in Subsection~{\ref{subsec:decoding}}, after which we discuss the  training methodology in Subsection~\ref{subsec:training}. Complexity, latency, and  properties of the decoder are discussed  in Subsections~\ref{subsec:complexity}-\ref{subsec:discussion}, respectively.

\begin{figure}
\removelatexerror
  \begin{algorithm}[H]
    \caption{\acp{wbp} Ensemble Inference}
    \label{alg:ensemble}
    \SetAlgoLined
    \SetKwInOut{Input}{Input}
    \SetKwInOut{Output}{Output}
    \SetKwProg{TheEnsembleWBPAlgorithm}{Ensemble Inference}{}{}
    \Input{Received word $\boldsymbol{\ell}$ \newline \ac{crc} parity-check matrix $\boldsymbol{H}_{CRC}$ \newline \ac{bp} decoder $\mathcal{F}_0$ \newline Ensemble's decoders $\mathcal{F}_1,...,\mathcal{F}_\alpha$} 
    \Output{Decoded word $\hat{\boldsymbol{u}}^p$}    \TheEnsembleWBPAlgorithm{$(\boldsymbol{y},\boldsymbol{H}_{CRC},\mathcal{F}_0,\mathcal{F}_1,...,\mathcal{F}_\alpha)$}{
    
    \label{line:u0}
    Extract $\hat{\boldsymbol{u}}^{(0)}$ from $\hat{\boldsymbol{u}}^{p,(0)}= \mathcal{F}_0 (\boldsymbol{\ell})$\\ \label{line:extract_u0}
    
    $\boldsymbol{r}^{(0)} = \boldsymbol{H}_{CRC} \cdot \hat{\boldsymbol{u}}^{(0)}$\\ \label{line:r0}
    
    \uIf{$\boldsymbol{r}^{(0)} = \boldsymbol{0}$}
        {return $\hat{\boldsymbol{u}}^{p,(0)}$\\ \label{line:return0}} 
        
    \For{$i\in\{1,...,\alpha\}$}{ 
    
            \label{line:ui}
            Extract $\hat{\boldsymbol{u}}^{(i)}$ from $\hat{\boldsymbol{u}}^{p,(i)} =  \mathcal{F}_i (\boldsymbol{\ell})$\\ \label{line:extract_ui}
            
            $\boldsymbol{r}^{(i)} = \boldsymbol{H}_{CRC} \cdot \hat{\boldsymbol{u}}^{(i)}$\\ \label{line:ri}
            
        }
    \uIf{$\exists i\in \{1,...,\alpha\}$  such that $\boldsymbol{r}^{(i)}=\boldsymbol{0}$} 
        {return $\hat{\boldsymbol{u}}^{p,(i)}$\\ \label{line:return1}} 
        
    Find $j$ such that $\boldsymbol{r}^{(0)}\in \mathcal{\bar{R}}^{(j)}$ \\ \label{line:j_mapping}
    
    \KwRet{$\hat{\boldsymbol{u}}^{p,(j)}$} \label{line:return3}
  }
  \end{algorithm}
\end{figure}

\vspace{-0.2cm}
\subsection{Inference Algorithm}
\label{subsec:decoding}

Our proposed decoder operates in two stages, designed to facilitate low-latency decoding by building upon the insight that in many cases, standard \ac{bp} decoding is sufficient. We employ a gated ensemble architecture, which is utilized only when standard \ac{bp} fails, leveraging the presence of \ac{crc} codes to identify such errors. Accordingly, the gating function $\mathcal{F}_0$ is a \ac{bp} decoder that recovers low-noise channel words with only small complexity overhead. 
Words that are incorrectly decoded are  passed to the ensemble, composed of multiple trainable decoders $\mathcal{F}_{\rm WBP}$, for further processing. 
Let us denote each $\mathcal{F}_{\rm WBP}$ member of the ensemble with a different index $\mathcal{F}_i$ ($i\geq 1$), and the number of members as $\alpha$. Thus, the set of decoders in the ensemble is $\{\mathcal{F}_1,\ldots, \mathcal{F}_\alpha\}$.

\begin{figure}
    \centering
    \includegraphics[width=\columnwidth]{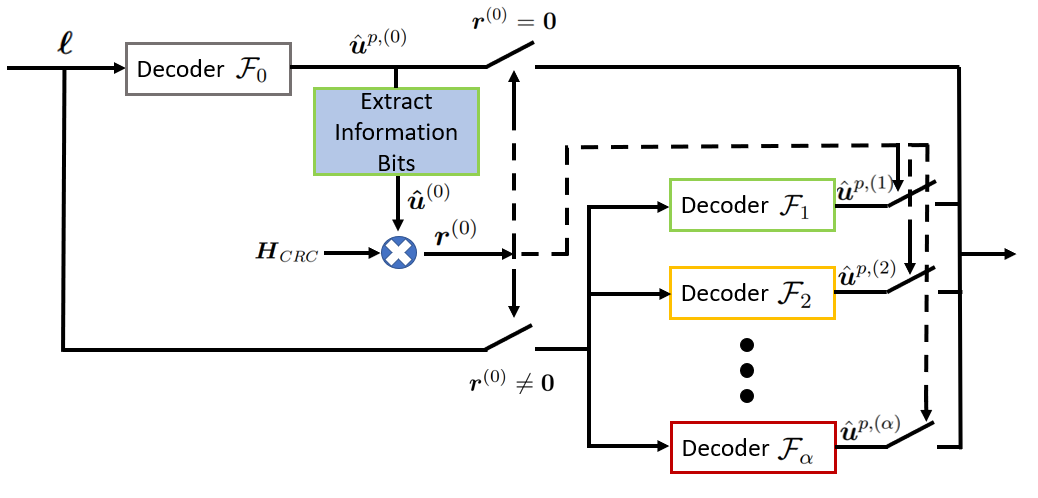}
    \caption{CRC-based Ensemble of \ac{wbp} Decoders}
    \label{fig:ensemble_decoder}
\end{figure}

Each word that enters the ensemble is decoded by every \ac{wbp} member $\mathcal{F}_i$ that outputs $\hat{\boldsymbol{u}}^{p,(i)}$. From each $\hat{\boldsymbol{u}}^{p,(i)}$, the estimated \ac{crc} word $\hat{\boldsymbol{u}}^{(i)}$ is extracted, and the ensemble's decision is based on the different outputs. If any member of the ensemble was correct (i.e.,  $\boldsymbol{r}^{(i)} = \boldsymbol{0}$), its decision $\hat{\boldsymbol{u}}^{p,(i)}$ is outputted. If no validation rule applies, we match the \acp{msb} of $\boldsymbol{r}^{(0)}$ to a region $\mathcal{\bar{R}}^{(j)}$ (explained in  Subsection~\ref{subsec:training}) and output the corresponding word $\hat{\boldsymbol{u}}^{p,(j)}$. The procedure is summarized as Algorithm~\ref{alg:ensemble} and illustrated in Fig.~\ref{fig:ensemble_decoder}.

\vspace{-0.2cm}
\subsection{Training Method}
\label{subsec:training}


To form each ensemble member $\mathcal{F}_i$ we train each one on different dataset $\mathcal{D}^{(i)}$. The ensemble should not be trained end-to-end, as we want diverse members trained on different data sets. 
We divide the dataset $\mathcal{D}$ into multiple different datasets $\{\mathcal{D}^{(i)}\}_{i=1}^{\alpha}$ based on the \acp{msb} of the \ac{crc} (assuming $\log_2 \alpha$ is an integer). Specifically, we define the set $\mathcal{\bar{R}}^{(i)}$ by considering the $\log_2 \alpha$ \acp{msb} of  the remainder $\boldsymbol{r}$ \eqref{eqn:parity-check}:
\begin{equation} \label{eq:crc_partition}
    \mathcal{\bar{R}}^{(i)}=\Big\{\boldsymbol{r}: \sum_{j=0}^{\log_2 \alpha -1} r_j 2^{j} = i -1;\, \boldsymbol{r}\neq \boldsymbol{0}\Big\}.
\end{equation}

By definition, the sets $\{\mathcal{\bar{R}}^{(i)}\}_{i=1}^{\alpha}$ are distinct and cover all possible error patterns. 
Based on this partition, we obtain the data $\{\mathcal{D}^{(i)} \}_{i=1}^{\alpha}$ for training by simulating $K$ labeled message words $\{\boldsymbol{u}^{p}_k\}$, and transmitting them through the channel to get $\{\boldsymbol{\ell}_k\}$. The resulting datasets are given by: 
\begin{equation}
\label{eq:datasets_induction}
    \mathcal{D}^{(i)} = \{(\boldsymbol{\ell}_k,\boldsymbol{u}^{p})_k:\boldsymbol{r}_k^{(0)} =  \boldsymbol{H}_{CRC} \cdot \hat{\boldsymbol{u}}^{(0)}_k\in \mathcal{\bar{R}}^{(i)}\}.
\end{equation}
Then, the $i^\text{th}$ member is trained using dataset $\mathcal{D}^{(i)}$, following the  procedure explained in Subsection~\ref{subsec:wbp} with the \ac{bce} loss.

\vspace{-0.4cm}
\subsection{Complexity and Latency Analysis}
\label{subsec:complexity}

Our proposed decoder is designed to operate in hardware-constrained real-time systems; therefore we next analyze its  complexity (i.e., number of parameters) and inference latency.

{\bf Complexity:}
Consider a trained decoder $\mathcal{F}_i$ running for a maximum of $T$ iterations. For the unfolded polar \ac{wbp} decoder in Subsection~\ref{subsec:wbp}, the number of weights for a polar encoder with codeword size $N_c$ equals 
  $4T\log{N_c}$ .
 Thus, an ensemble with $\alpha$ \ac{wbp} decoders, and the vanilla gating decoder $\mathcal{F}_0$, has a total of $(\alpha+1) \cdot 4T\log{N_c}$ weights.

 \begin{figure*}
    \centering
    \begin{subfigure}[a]{0.32\textwidth}  
        \centering 
        \includegraphics[width=\textwidth]{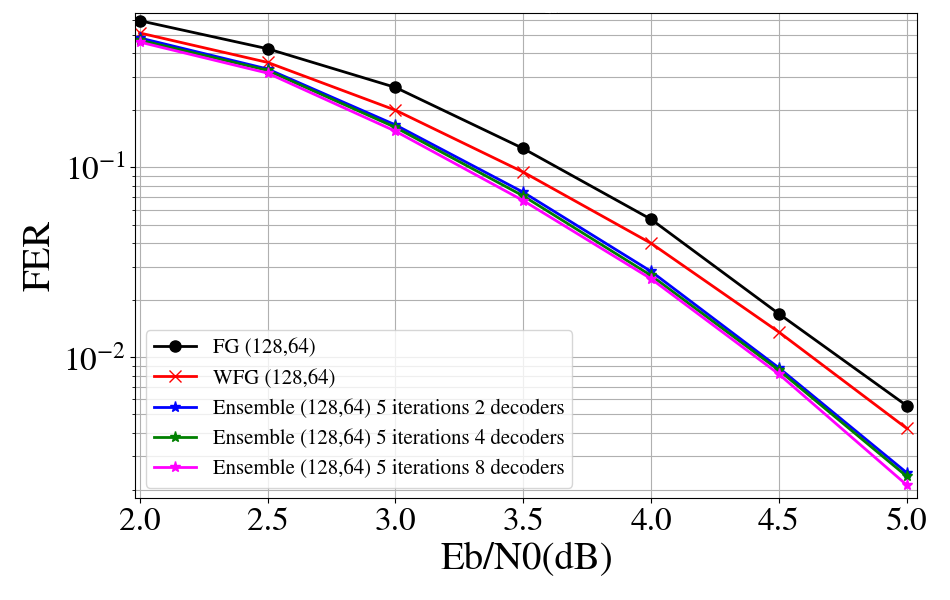}
        \caption{$(128,64)$ polar code}
    \end{subfigure}
    \hfill
    \begin{subfigure}[a]{0.32\textwidth}  
        \centering 
        \includegraphics[width=\textwidth]{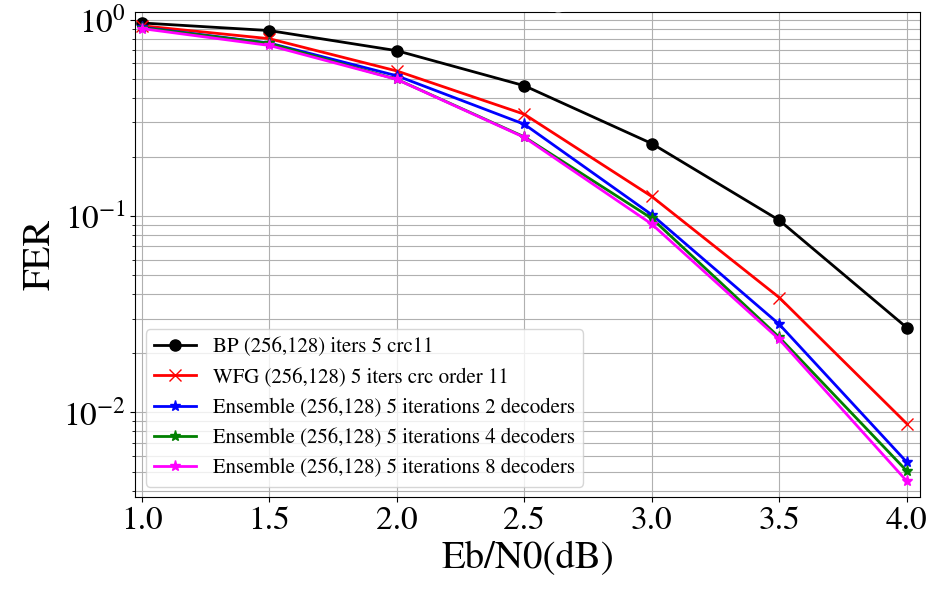}
        \caption{$(256,128)$ polar code}
    \end{subfigure}
        \hfill
    \begin{subfigure}[a]{0.32\textwidth}  
        \centering 
        \includegraphics[width=\textwidth]{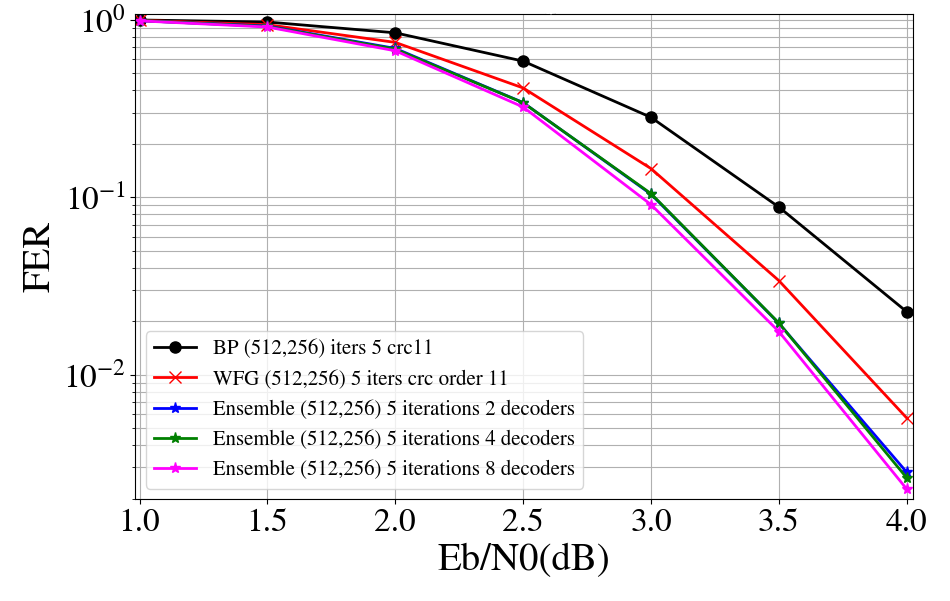}
        \caption{$(512,256)$ polar code}
    \end{subfigure}
    \caption{\ac{fer} performance for different polar codes}
    \label{fig:FER_performance}
    \vspace{-0.4cm}
\end{figure*}


{\bf Latency:} In principle, one can measure latency via run-time simulation. However, such results are usually highly dependent on the specific hardware that simulates the setup. As such, we adopt a complexity-based latency measure, through the number of \ac{flops}, and specifically the number of multiplications, which faithfully reflect on run-time in a hardware-invariant manner.

For a single model, assuming that \ac{crc} check has a negligible latency, the overall latency measure $\tau_i$ is dictated by the weights, i.e., 
   $ \tau_i = 4T\log{N_c}$.  
However, in order to minimize the excessive latency, the ensemble is triggered in case that the vanilla decoder $\mathcal{F}_0$ failed to decode correctly. Moreover, each member is applied independently and in parallel of the other models. Consequently, the average latency is given by
\begin{equation} \label{eq:parallel_runtime}
    \tau_{\rm ensemble} = (1 + \Pr(\boldsymbol{r}^{(0)} \neq \boldsymbol{0}))\cdot 4T\log{N_c},
\end{equation}
where $\Pr(\boldsymbol{r}^{(0)} \neq \boldsymbol{0})$ is the probability that the gating \ac{bp} decoder fails. This gated operation implies that \eqref{eq:parallel_runtime} yields only minor excessive latency over a single \ac{bp} decoder, 
as  demonstrated in Section~\ref{sec:numerical_evaluations}.



\vspace{-0.4cm}
\subsection{Discussion}
\label{subsec:discussion}
The  proposed ensemble decoder jointly accounts for \ac{crc} and polar codes, which are both utilized in 5G-NR specifications, making it naturally applicable with existing standards. Moreover, it combines  two emerging decoding paradigms: list-decoding and neural-decoding, aiming to combine the best of the two. We employ \ac{crc} in novel and unseen ways: (1) to form the ensemble's experts, and (2) to select the most probable estimated word. 
Our complexity and latency analysis indicate that 
the scheme is practical in the sense of both software and hardware sides, inducing low complexity overhead. 

The ensembling algorithm can be further improved by increasing diversity in the data or the architecture sides. On the data side, one can combine this partition with previous partitions \cite{raviv2020data} that incorporate  information other than \ac{crc} bits, e.g., by the codeword Hamming distance or by the estimated errors. On the architecture side, it could be enhanced by combining different base-experts, such as full neural network decoders (at the possible cost of limiting their applicability to large blocks), or the Tanner-based decoder. These extensions 
are left for future study.
	\vspace{-0.2cm}
	\section{Numerical Evaluations}
\label{sec:numerical_evaluations}
\vspace{-0.1cm} 
In this section we numerically evaluate the proposed ensemble scheme over a number of scenarios to prove its gains. 

\smallskip
{\bf Simulation Setup:}
We simulate\footnote{The source code used in our experiments is available at \href{https://github.com/tomerraviv95/polar-ensembles}{https://github.com/tomerraviv95/polar-ensembles}} an \ac{awgn} channel as described in \ref{subsec:problem_formulation}. 
we refer to the \ac{snr} as the normalized \ac{snr} ($E_b / N_0$), commonly used in digital communication. We consider $T=5$ iterations for \ac{bp} and \ac{wbp} decoding as the common benchmark. Training is done on the zeros codeword $\boldsymbol{0}$ only, due to the symmetry of the \ac{bp} (see \cite{nachmani2018deep} for further details), with a learning rate of ${10}^{-2}$.

To form the ensemble's experts, we train each expert by simulating words  from \ac{snr} values of $\{2,3,4,5\}$ in dB for the $(128,64)$ polar code; and \ac{snr} of $\{1,2,3,4\}$ in dB for polar codes of $(256,128)$ or $(512,256)$. For each \ac{snr}, we generated $10^5\cdot\alpha$  words, and divided them across the different $\mathcal{D}^{(i)}$ via \eqref{eq:datasets_induction}. These \ac{snr} values neither have too noisy words nor too many correct words thus contain words that are useful for training. Every $\mathcal{F}_i$ member is trained on for $100$ epochs of $200$ batches, each of size $|\mathcal{D}^{(i)}|/200$, using the \ac{bce} loss. 

\begin{figure*}
    \begin{subfigure}[b]{0.32\textwidth}   
        \centering
        \includegraphics[width=\columnwidth]{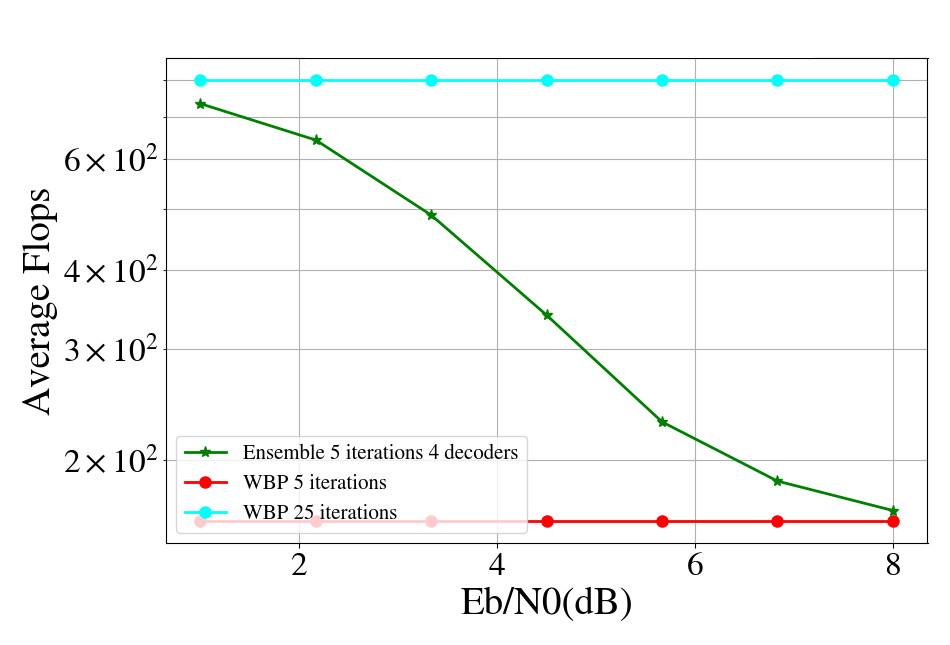}
        \caption{Polar code $(256,128)$}
        \label{fig:flops_256_128}
    \end{subfigure}
    \hfill
        \begin{subfigure}[b]{0.32\textwidth}   
        \centering
        \includegraphics[width=\columnwidth]{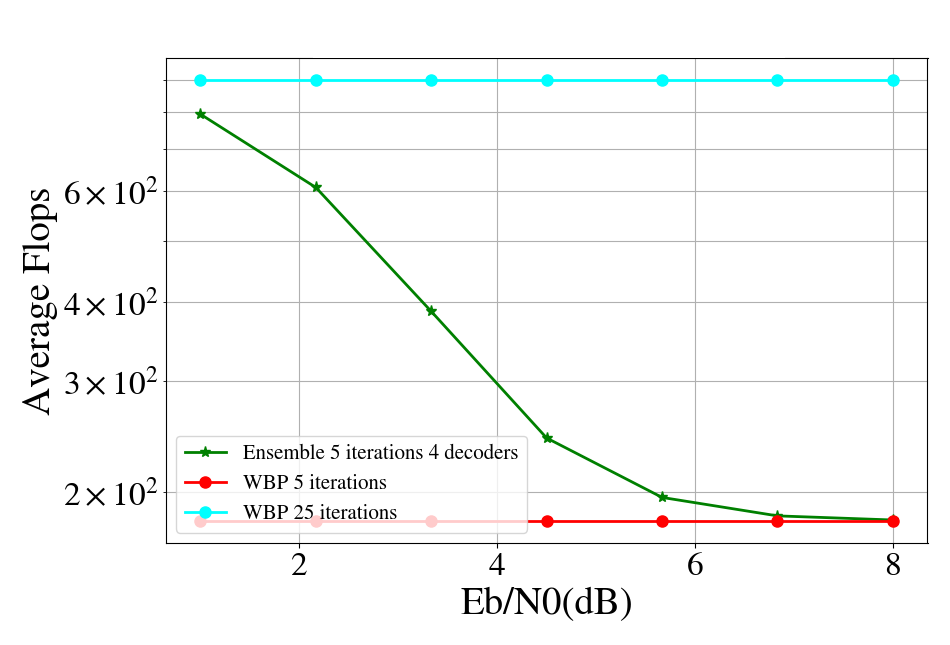}
        \caption{Polar code $(512,256)$}
        \label{fig:flops_512_256}
    \end{subfigure}
    \hfill
    \begin{subfigure}[b]{0.32\textwidth}   
        \centering
        \includegraphics[width=\columnwidth]{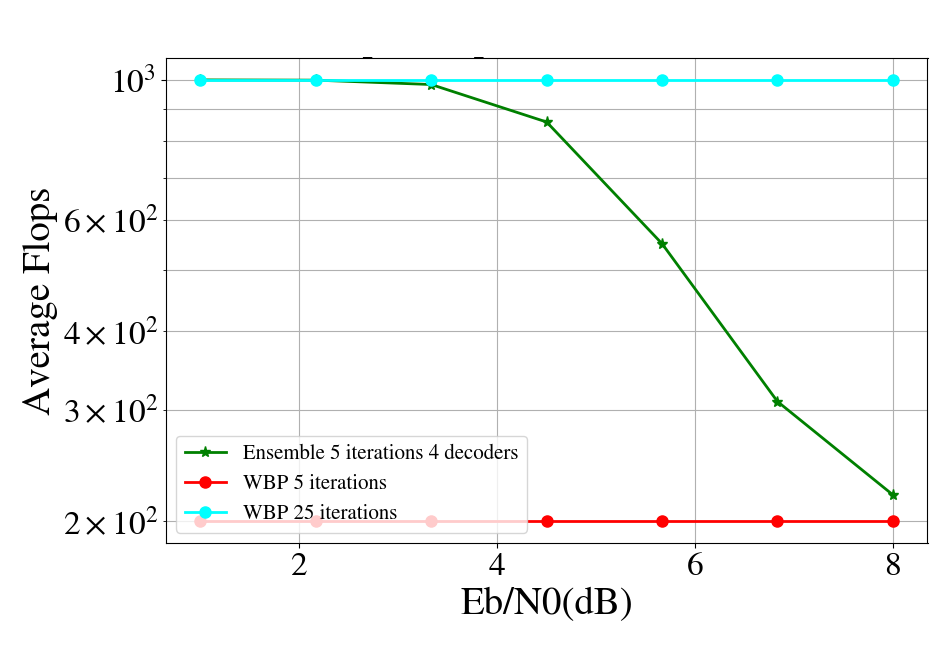}
        \caption{Polar code $(1024,512)$}
        \label{fig:flops_1024_512}
    \end{subfigure}
    \caption{Average number of flops by the \ac{snr} value, $\alpha = 4$} 
    \label{fig:flops}
    \vspace{-0.4cm}
\end{figure*}

\smallskip
{\bf \ac{fer} Results:} 
We first evaluate performance in terms of \ac{fer}. Here, the validation dataset is composed of \ac{snr} values of $1$dB to $4$dB, spaced by $0.5$dB, and at each point at least 500 errors are accumulated. Evaluation of the scheme is done by measuring the \ac{fer} metric. The results for polar codes of sizes $(128,64)$, $(256,128)$ and $(512,256)$  using ensembles with $\alpha \in \{2,4,8\}$ members are presented in Fig.~\ref{fig:FER_performance}. We observe in Fig.~\ref{fig:FER_performance} that as the number of decoders increases, the \ac{fer} decreases, however in negligible gains. By using only 2 decoders, the ensemble achieves gains of about $0.25$dB in \ac{fer} for the three codes in medium-to-high \acp{snr}. Note that gains of around $0.1$dB are achieved in the low \acp{snr} regime, which is shown to be harder for improvement using neural approaches (see \cite{nachmani2018deep}). Raising the number of members to $\alpha=8$ yields additional $0.05$dB for high length codes. As evident from the above simulations, our approach achieves consistent gains even at high code lengths where other  methods \cite{choukroun2022error,nachmani2020gated} are infeasible.

\smallskip
{\bf Latency Analysis:}
As described in Subsection~\ref{subsec:complexity}, the complexity of the model can be estimated by the number of weights used in run-time, which can in turn be used to derive the total latency. Fig.~\ref{fig:flops} shows the latency evaluation according to the analysis in~\eqref{eq:parallel_runtime}. We first plot the lower bound on the number of \ac{flops}, i.e., the latency of $\mathcal{F}_0$, as well as the upper bound, which is the latency of $(\alpha+1)$ decoders.
We observe in Fig.~\ref{fig:flops_256_128} that for  a medium length code, the complexity drops quickly as the \ac{snr} increases. As the code length increases, as in 
 Fig.~\ref{fig:flops_512_256} and Fig.~\ref{fig:flops_1024_512}, the drop in \ac{flops} as a function of \ac{snr} becomes more steep but starts in higher \acp{snr} values. This results from the  term $\Pr(\boldsymbol{r}^{(0)} \neq \boldsymbol{0})$.

	\vspace{-0.2cm}
	\section{Conclusion}
	\label{sec:conclusion}
	\vspace{-0.1cm}
We introduced a novel polar decoder which combines list-decoding and neural-decoding via gated ensembles. 
We have shown that moderate gains of around $0.25$dB may be achieved at different code lengths using this hybrid approach, at a manageable complexity overhead. Our approach can be applied to long polar codes, where many  neural-decoding methods still struggle.
	\vspace{-0.2cm}
	\bibliographystyle{IEEEtran}
	\bibliography{IEEEabrv,refs}
\end{document}